\documentstyle[]{mn}
\newif\ifAMStwofonts
\def\ee #1 {\times 10^{#1}}
\def\ut #1 #2 { \, \rmn{#1}^{#2}}
\def\u #1 { \, \rmn{#1}}

\def\persec {\, \hbox{s}^{-1}}
\def\percc {\,\rmn{cm}^{-3}}
\def\micron {\, \mu \hbox{m}}

\let\grad=\nabla
\def\cross{\bmath{\times}}
\def\curl #1 {\grad \cross #1}
\def\div #1 {\grad \cdot #1}
\def\nh{{n_{\rm H}}}

\def\nh{{n_{\rm H}}}

\def\v{\bmath{v}}

\def\x{\bmath{x}}

\def\B{\bmath{B}}
\def\Bh{\bmath{\hat{B}}}
\def\E{\bmath{E}}            % E
\def\Epa{\bmath{E'_\parallel}}  % E'_||
\def\Epe{\bmath{E'_\perp}}  % E'_perp
\def\J{\bmath{J}}

\def\dv{\bmath{\delta\v}}
\def\dE{\bmath{\delta\E}}
\def\dB{\bmath{\delta\B}}
\def\sigv{<\sigma v>}

\newcommand{\pddt} [1] {\frac{\partial #1}{\partial t}}
\ifoldfss
  \newcommand{\rmn}[1] {{\rm #1}}

  \ifCUPmtlplainloaded \else
    \NewTextAlphabet{textbfit} {cmbxti10} {}
    \NewTextAlphabet{textbfss} {cmssbx10} {}
    \NewMathAlphabet{mathbfit} {cmbxti10} {} % for math mode
    \NewMathAlphabet{mathbfss} {cmssbx10} {} %  "   "    "
  \fi
  \ifAMStwofonts
    \ifCUPmtlplainloaded \else
      \NewSymbolFont{upmath} {eurm10}
      \NewSymbolFont{AMSa} {msam10}
      \NewMathSymbol{\upi}     {0}{upmath}{19}
      \NewMathSymbol{\umu}     {0}{upmath}{16}
      \NewMathSymbol{\upartial}{0}{upmath}{40}
      \NewMathSymbol{\leqslant}{3}{AMSa}{36}
      \NewMathSymbol{\geqslant}{3}{AMSa}{3E}

    \fi
  \fi
\fi % End of OFSS
\ifnfssone
  \newmathalphabet{\mathit}
  \addtoversion{normal}{\mathit}{cmr}{m}{it}
  \addtoversion{bold}{\mathit}{cmr}{bx}{it}
  \newcommand{\rmn}[1] {\mathrm{#1}}

  \newmathalphabet{\mathbfit} % math mode version of \textbfit{..}
  \addtoversion{normal}{\mathbfit}{cmr}{bx}{it}
  \addtoversion{bold}{\mathbfit}{cmr}{bx}{it}
  \newmathalphabet{\mathbfss} % math mode version of \textbfss{..}
  \addtoversion{normal}{\mathbfss}{cmss}{bx}{n}
  \addtoversion{bold}{\mathbfss}{cmss}{bx}{n}
  \ifAMStwofonts
    \ifCUPmtlplainloaded \else
      \UseAMStwoboldmath
      \makeatletter
      \new@mathgroup\upmath@group
      \define@mathgroup\mv@normal\upmath@group{eur}{m}{n}
      \define@mathgroup\mv@bold\upmath@group{eur}{b}{n}
      \edef\UPM{\hexnumber\upmath@group}
      \new@mathgroup\amsa@group
      \define@mathgroup\mv@normal\amsa@group{msa}{m}{n}
      \define@mathgroup\mv@bold\amsa@group{msa}{m}{n}
      \edef\AMSa{\hexnumber\amsa@group}
      \makeatother
      \mathchardef\upi="0\UPM19
      \mathchardef\umu="0\UPM16
      \mathchardef\upartial="0\UPM40
      \mathchardef\leqslant="3\AMSa36
      \mathchardef\geqslant="3\AMSa3E
    \fi
  \fi
\fi % End of NFSS release 1
\ifnfsstwo
  \newcommand{\rmn}[1] {\mathrm{#1}}

  \DeclareMathAlphabet{\mathbfit}{OT1}{cmr}{bx}{it}
  \SetMathAlphabet\mathbfit{bold}{OT1}{cmr}{bx}{it}
  \DeclareMathAlphabet{\mathbfss}{OT1}{cmss}{bx}{n}
  \SetMathAlphabet\mathbfss{bold}{OT1}{cmss}{bx}{n}
  \ifAMStwofonts
    \ifCUPmtlplainloaded \else
      \DeclareSymbolFont{UPM}{U}{eur}{m}{n}
      \SetSymbolFont{UPM}{bold}{U}{eur}{b}{n}
      \DeclareSymbolFont{AMSa}{U}{msa}{m}{n}
      \DeclareMathSymbol{\upi}{0}{UPM}{"19}
      \DeclareMathSymbol{\umu}{0}{UPM}{"16}
      \DeclareMathSymbol{\upartial}{0}{UPM}{"40}
      \DeclareMathSymbol{\leqslant}{3}{AMSa}{"36}
      \DeclareMathSymbol{\geqslant}{3}{AMSa}{"3E}
    \fi
  \fi
\fi % End of NFSS release 2
\ifCUPmtlplainloaded \else
  \ifAMStwofonts \else % If no AMS fonts
    \def\upi{\pi}
    \def\umu{\mu}
    \def\upartial{\partial}
  \fi
\fi
\input epsf
\title[Conductivity of molecular gas]
{The conductivity of dense molecular gas}
\author[M. Wardle and C. Ng]
       {Mark Wardle$ ^1 $ \& Cindy Ng$ ^2 $ \\
$ ^1 $Special Research Centre for Theoretical Astrophysics, University
of Sydney, NSW 2006, Australia \\
$ ^2 $Department of Physics, University of Tasmania, Hobart, 
TAS 7001, Australia}
\date{MNRAS accepted 1998 October 21}
\pagerange{\pageref{firstpage}--\pageref{lastpage}}
\pubyear{1998}
\begin{document}
\maketitle
\label{firstpage}
\begin{abstract}
We evaluate the conductivity tensor for molecular gas at densities 
ranging from $ 10^4 $ to $ 10^{15} \percc $ for a variety of grain 
models.  The Hall contribution to the conductivity has generally been neglected in 
treatments of the dynamics of molecular gas.  We find that it is not 
important if only 0.1 $ \micron $ grains are considered, but for a 
Mathis-Rumpl-Nordsieck grain-size distribution (with or without PAHs) 
it becomes important for densities between $ 10^7 $ and $ 10^{11} 
\percc $.  If PAHs are included, this range is reduced to $ 10^9 
$--$ 10^{10} \percc $.

The consequences for the magnetic field evolution and dynamics of 
dense molecular gas are profound.  To illustrate this, we consider the 
propagation of Alfv\'en waves under these conditions.  A linear 
analysis yields a dispersion relation valid for frequencies below the 
neutral collision frequencies of the charged species.  The dispersion 
relation shows that there is a pair of circularly polarised modes with 
distinct propagation speeds and damping rates.  We note that the 
gravitational collapse of dense cloud cores may be substantially modified 
by the Hall term.

\end{abstract}
\begin{keywords}
magnetohydrodynamics -- Alfv\'en waves -- dust grains -- molecular 
clouds.
\end{keywords}

\section{Introduction}

The breakdown of flux-freezing in molecular clouds has generally been 
treated using the concept of ``ambipolar diffusion'' (e.g.  Spitzer 
1978).  In this approximation the magnetic field is regarded as being 
frozen into the \emph{ionised} component of the fluid, which drifts as 
a whole with respect to the neutrals.  During ambipolar diffusion, the 
magnetic stresses operating on the ionised plasma are transmitted to 
the dominant neutral component via collisions.  The resulting 
dissipation damps MHD waves (Kulsrud \& Pearce 1969) and is 
particularly important in shock waves where it provides the heating 
within the shock front (Draine 1980).  Ambipolar diffusion allows the 
trickling of neutrals through the ions and magnetic field towards the 
centre of a dense core until the core becomes gravitationally unstable 
and collapses (see e.g. Mouschovias 1987; Fiedler \& Mouschovias 1993).

The ambipolar-diffusion approximation is valid when the magnetic force 
on each charged particle dominates the drag force arising through 
collisions with the neutrals (i.e.  the product of the gyrofrequency 
and the time scale for momentum-exchange with the neutrals is much 
greater than unity).  This is generally an excellent approximation for 
ions and electrons in molecular clouds, except for densities and 
magnetic field strengths appropriate for protostellar disks (Norman \& 
Heyvaerts 1985; Wardle \& K\"onigl 1993).

Charged grains, however, are partially decoupled from the magnetic 
field by neutral collisions because of their large geometrical 
cross-section.  This effect has generally been treated by including a 
separate equation of motion for the grains so that their drift speed 
through the neutrals can be calculated (see e.g.  Draine 1980).  The 
current density then has a component out of the plane containing the 
magnetic field and the electric field in the rest frame of the neutral 
fluid.  This component, the Hall current, has generally been suppressed 
in calculations of gravitational condensation via ambipolar diffusion 
(Ciolek \& Mouschovias 1993), and in shock modelling (e.g.  Draine, 
Roberge \& Dalgarno 1983; Wardle \& Draine 1987; Kaufman \& Neufeld 
1996).  This is a poor approximation when the grain drag is important 
(typically at densities in excess of $ 10^6 \percc $.  In the case 
of shock waves, when this component is retained the shock structure is 
significantly thinner (Pilipp, Hartquist \& Havnes 1990) and is 
qualitatively different, exhibiting a twist back and forth around the 
preshock normal (Wardle 1998).

Early studies assumed that grains were of a single characteristic size 
(typically $ 0.1 \micron $).  It has since been recognised that the MRN 
grain-size distribution (Mathis, Rumpl \& Nordsieck 1977) and the presence 
of polycyclic aromatic hydrocarbons (PAHs) (Leger \& Puget 1984) 
imply that grains play a more important role in coupling the 
magnetic field to the neutral gas (Nishi, Nakano \& Umebayashi 1991), 
grains and PAHs being the dominant charged species above $ 10^7 
\percc $.

Here we examine the conductivity of molecular gas at densities 
relevant to cloud cores, and at higher densities occurring during the 
formation of a protostar and its associated disk.  We formulate MHD in 
\S\ref{sec:formulation} in terms of a conductivity tensor appropriate 
for a weakly-ionised gas (Cowling 1957; Norman \& Heyvaerts 1985; 
Nakano \& Umebayashi 1986) and stress that the \emph{vector} evolution 
of the magnetic field depends on the relative magnitudes of the 
different components.  We evaluate the conductivity for single-size, 
MRN, and MRN-plus-PAH grain models in \S \ref{sec:ionisation}, with 
charged particle abundances from Umebayashi \& Nakano (1990) and Nishi 
et al (1991).  We find that the Hall term is not particularly 
important in the single-size grain model, but in the other models 
becomes important for densities between $ 10^7 $ and $ 10^{11} 
\percc $.  The dynamical behaviour of molecular gas at higher 
densities is therefore profoundly different from the ambipolar 
diffusion case.  By way of illustration we consider the propagation of 
Alfv\'en waves in \S\ref{sec:Alfven_waves}.  The implications of our 
results are further discussed in \S\ref{sec:discussion}, and our 
conclusions are summarised in \S\ref{sec:summary}.

\section{Formulation}
\label{sec:formulation}
\subsection{MHD in weakly-ionised media}

Our definition of ``weakly-ionised'' is that the abundances of 
charged species are so low that their inertia and thermal pressure is 
negligible, and that the changes in the neutral gas resulting from 
ionisation and recombination are so small that they may also be 
neglected.

In this limit, the fluid equations may be written
\begin{equation}
\pddt{\rho} + \div(\rho \v) = 0 \,,	
	\label{eq:continuity}
\end{equation}
\begin{equation}
\rho\pddt{\v} + \rho(\v \cdot \grad)\v + c_s^2\grad\rho = 
\frac{\J\cross\B}{c} \,,	
	\label{eq:momentum}
\end{equation}
where $ c_s^2 $, the isothermal sound speed, is assumed to be  
constant,
\begin{equation}
\J = \frac{c}{4\pi}\grad\cross \B \,,
\label{eq:j_curlB}
\end{equation}
\begin{equation}
\pddt{\B} = \curl (\v \cross \B) - c \curl \E' \,,
\label{eq:induction}
\end{equation}
and
\begin{equation}
\div \B = 0 \,.
\label{eq:divB}
\end{equation}
Here
\begin{equation}
	\E'=\E+\v\cross\B/c
	\label{eq:efluid}
\end{equation}
is the electric field in the frame
comoving with the fluid, which is related to the 
current density by the conductivity tensor $ 
\bmath{\sigma} $:
\begin{equation}
	\J = \bmath{\sigma}\cdot \E'  \,.
	\label{eq:J-E}
\end{equation}

We shall consider an explicit expression for $ \bmath{\sigma} $ in 
the next subsection.  For now, we merely note that all of the 
information concerning the charged species is hidden in $ 
\bmath{\sigma} $, and that we are implicitly assuming a prescription 
for determining the abundances of the charged species so that the 
conductivity can be calculated.

At this point we emphasise the role that the conductivity plays in 
determining the evolution of the magnetic field through the induction 
equation (\ref{eq:induction}), which on substituting for $ \E' $ 
becomes:
\begin{equation}
\pddt{\B} = \curl (\v \cross \B) - 
\frac{c^2}{4\pi} \curl [ \bmath{\sigma^{-1}\cdot}\, (\curl \B) ] \,.
\label{eq:tensor_induction}
\end{equation}
The second term becomes important if the current associated with the 
gradient in the magnetic field is so large that $ |\E'| $ is 
comparable to or greater than $ |\v\cross\B|/c $.  The magnitude of 
the conductivity determines a characteristic length scale (or time 
scale) below which the field cannot be regarded as frozen-in (see e.g.  
Parker 1979).  It is not, however, the magnitude on which we wish to 
focus but rather the \emph{tensor} nature of the conductivity, which 
affects the \emph{direction} of $ \partial\B/\partial t $.  That is, 
the vector evolution of $ \B $ from an initial configuration depends 
on the relative magnitude of the components of $ \bsigma $.

\subsection{The conductivity tensor}

We characterise the various charged species in the neutral fluid by 
particle mass $m_j$ and charge $Z_je$, number density 
$n_j$, and drift velocity through the neutral gas $\v_j$, where 
the subscript $ j $ denotes different species.
Note that we have already implicitly assumed charge neutrality, i.e.
\begin{equation}
\sum_j n_j Z_j = 0 \,.
\label{eq:charge_neutrality}
\end{equation}
Assuming that the fluid evolves on a time scale that is long compared 
to the collision time scale of any type of charged particle with the 
neutrals, each charged particle drifts through the neutrals at a rate 
and direction determined by the instantaneous Lorentz force on the 
particle:
\begin{equation}
Z_j e(\E' + {\v_j \over c} \cross \B) - \gamma_j m_j \v_j
= 0 \,,
\label{eq:charged_drift}
\end{equation}
where the third term represents the drag force contributed by 
collisions with the neutrals, and
\begin{equation}
	\gamma_j = \frac{<\sigma v>_j}{m_j+m}
	\label{eq:gamma_j}
\end{equation}
where $ <\sigma v>_j $ is the rate coefficient for momentum transfer 
by collisions with the neutrals and $ m $ is the mean neutral 
particle mass.
The Hall parameter for species j,
\begin{equation}
\beta_j= {Z_jeB \over m_j c} \, {1 \over \gamma_j \rho}\, ,
\label{eq:Hall_parameter}
\end{equation}
determines the relative importance of the Lorentz and drag forces in 
balancing the electric force.

\begin{figure}
\centerline{\epsfxsize=8cm \epsfbox{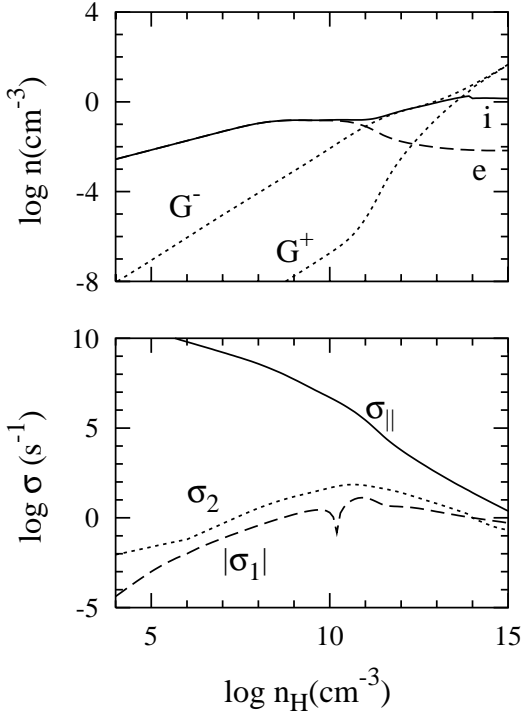}}\vskip 0cm 
\caption{\emph{Upper panel:} the abundances of charged species when 
grains have size $ 0.1 \micron $.  The curves labelled $ i $, $ e 
$, $ G^+ $ and $ G^- $ refer to ions, electrons and positively 
and negatively charged grains respectively (adapted from Umebayashi \& 
Nakano 1990).  \emph{Lower panel:} the field-parallel ($ 
\sigma_\parallel $), Hall ($ \sigma_1 $) and Pedersen ($ \sigma_2 
$) components of the conductivity tensor.  The ambipolar-diffusion 
approximation is valid for $ |\sigma_1| \ll \sigma_2 \ll 
\sigma_\parallel $.  The spike in the curve for $| \sigma_1 |$ at 
$ \nh\approx 10^{10} \percc $ occurs because $ \sigma_1 $ is 
negative at lower densities and positive at higher densities.}
\label{fig:NU90_conductivity} %\subsubsection{fig:NU90 conductivity}
\end{figure}

\begin{figure}
\centerline{\epsfxsize=8cm \epsfbox{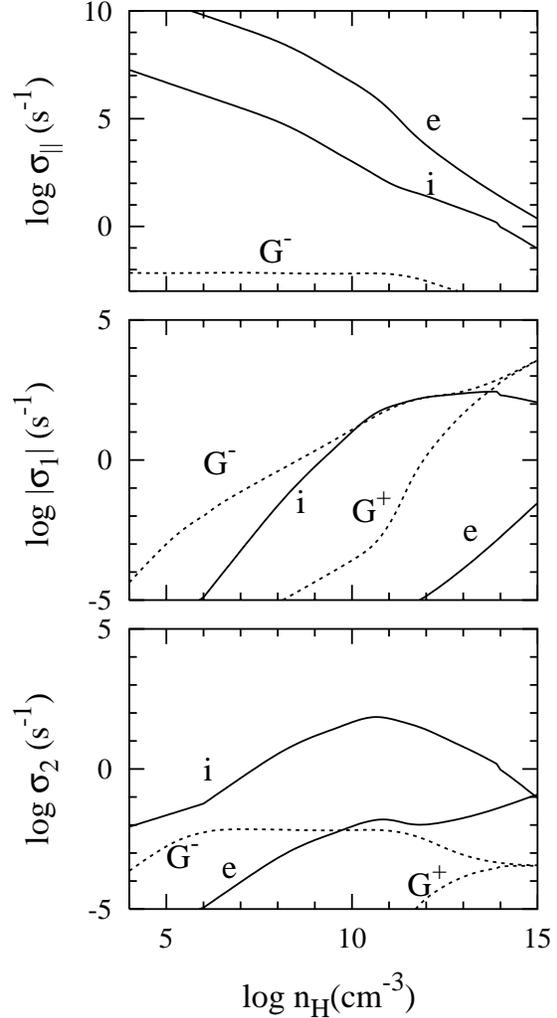}}\vskip 0cm \caption{The 
contributions made by different species to the components of the 
conductivity tensor plotted in Fig.  \ref{fig:NU90_conductivity}.  
Note that the positive and negative species contribute positive and 
negative terms to $ \sigma_1 $ respectively, but that all species 
contribute positively to $ \sigma_\parallel$ and $\sigma_2$.}
\label{fig:NU90_contributions} %\subsubsection{fig:NU90 contributions}
\end{figure}

Eq (\ref{eq:charged_drift}) can be inverted for 
$ \v_j $, which can then be used to form an expression for 
$ \J = e\sum_j n_j Z_j \v_j$:
\begin{equation}
	\J = \sigma_{\parallel} \Epa + \sigma_1 \Bh \cross \Epe + 
	\sigma_2 \Epe
	\label{eq:jsigmaE}
\end{equation}
where $ \Epa $ and $ \Epe $ are the decomposition of $ \E' $ 
into vectors parallel and perpendicular to $ B $ respectively.
The components of $ \bsigma $ are the 
conductivity parallel to the magnetic field,
\begin{equation}
	\sigma_{\parallel} = \frac{ec}{B}\sum_{j} n_j Z_j \beta_j \,,
	\label{eq:sigma0}
\end{equation}
the Hall conductivity,
\begin{equation}
	\sigma_1 = \frac{ec}{B}\sum_{j}\frac{n_j Z_j}{1+\beta_j^2}\,,
	\label{eq:sigma1}
\end{equation}
and the Pedersen conductivity
\begin{equation}
	\sigma_2 = \frac{ec}{B}\sum_{j}\frac{n_j Z_j \beta_j}{1+\beta_j^2}
	\label{eq:sigma2}
\end{equation}
(Cowling 1957; Norman \& Heyvaerts 1985; Nakano \& Umebayashi 1986).  
We shall find it useful to refer to the total conductivity 
perpendicular to the field,
\begin{equation}
	\sigma_{\perp} = \sqrt{\sigma_1^2 + \sigma_2^2} \,.
	\label{eq:sigma_perp}
\end{equation}

Three conductivity regimes are delineated by the magnitude of the 
typical Hall parameter, $ \beta $, of the charged species:
\begin{enumerate}
	\item  $ |\beta|\gg 1 $, which implies that $ \sigma_\parallel \gg 
	\sigma_2 \gg |\sigma_1| $ and gives rise to the ambipolar diffusion 
	regime;

	\item  $ |\beta| \ll 1 $, which yields $ \sigma_\parallel 
	\approx \sigma_2 \gg |\sigma_1| $, implying that the conductivity is scalar
	($ \J \approx \sigma_\parallel \E' $) --  the resistive regime; and 

	\item  $ |\beta| \approx 1 $, the Hall regime.
\end{enumerate}
The latter has been largely neglected, but we shall see that 
the grain-size distribution implies that it is relevant over 
a broad range of conditions in dense molecular gas.

\section{Conductivity in dense molecular gas}
\label{sec:ionisation}

The conductivity tensor depends on the 
abundances of charged species, which in turn depend on the choice of 
grain model as recombination 
generally occurs on the surface of grains (Nishi et al 1991).   We adopt three grain 
models:
\begin{enumerate}
	\item Single-size grains, characterised by a grain radius $ 0.1 
	\micron $ and a total grain mass that is one percent of the total 
	mass in hydrogen.

	\item An MRN model with a power-law distribution of grain sizes 
	between 50 and 2500 \AA:
	\begin{equation}
         \frac{dn}{da} = A\nh a^{-3.5} \,,
    \label{eq:mrn}
    \end{equation}
    where $n(a)$ is the number density of grains with radii smaller than 
    $a$, and $A=1.5\ee -25 \ut cm -2.5 $ (Draine \& Lee 1984).

	\item An MRN model with an additional component of very small 
	grains with $ a \sim 3 $ \AA , representing PAHs with an 
	abundance $ n_g / \nh = 2\ee -7 $.
\end{enumerate}

To evaluate the conductivity, we obtain the abundances of charged 
species from Fig.  2 of Umebayashi \& Nakano (1990) and Figs 1 and 4 
of Nishi et al (1991).  The species we include are molecular and metal 
ions i, electrons e, MRN grains (G$ ^+ $, G$ ^- $) and PAHs (g$ 
^+ $, g$ ^- $).  The rate coefficients for momentum transfer by 
elastic scattering of species $ j $ with neutrals, $ \sigv_j $ are 
given by
\begin{equation}
	\sigv_i = \sigv_g = 1.6 \ee -9 \ut cm 3 \ut s -1 \,,
	\label{eq:sigv_i}
\end{equation}
\begin{equation}
	\sigv_e = 1 \ee -15 \ut cm 2 \, \left(\frac{128kT_e}{9\pi m_e}\right)^{1/2}\,,
	\label{eq:sigv_e}
\end{equation}
and
\begin{equation}
	\sigv_G = \pi a^2 \, \left(\frac{128kT}{9\pi m}\right)^{1/2}\,,
	\label{eq:sigv_G}
\end{equation}
where $ T $ and $ T_e $ are the neutral and electron temperatures 
respectively.
The rate coefficient for the PAHs has been set equal to that for the 
ions as the induced-polarisation cross-section dominates the geometric 
cross-section for charged particles with sizes below $ 10 $ \AA.  The 
expressions for $ \sigv_G $ and $ \sigv_e $ are valid as long as the 
grain and electron drift speeds are less than the sound speeds in the 
neutrals and electrons respectively. 

To calculate the conductivities, we adopt $ T=T_e=30 \u K  $, a 
cosmic-ray ionisation rate per hydrogen nucleus 
$ \zeta = 10^{-17}\persec \ut H -1 $, 
$ m=2.33 m_p $ corresponding to a helium abundance of 
0.1 by number, and a magnetic field that follows the standard $ \nh^{1/2} $ 
scaling for densities below $ 10^6 \percc $ (Myers \& Goodman 1988) 
with a weaker dependence at higher densities:
\begin{equation}
	\left(\frac{B}{\u mG }\right) =  \left\{ 
	\begin{array}{l}
		\left(\nh / 10^6 \percc \right)^{1/2} \mathrm{if}\, \nh < 10^6 
		\percc ,
		\\ [5pt]
		\left(\nh / 10^6 \percc \right)^{1/4} \mathrm{otherwise.} 
	\end{array} \right.
	\label{eq:B_scaling}
\end{equation}
When integrating over the MRN size distribution we have assumed that 
the ratio of charged to neutral grains is independent of grain radius.

The upper panel of Figure \ref{fig:NU90_conductivity} shows the 
charged particle densities for the single-size grain model.  Ions and 
electrons are the dominant charged species for $ \nh\la 10^{10} 
\percc $, negatively-charged grains replace the electrons for $ 
\nh\ga10^{11}\percc $, and above $ 10^{14}\percc $ ions are 
dominated by positively charged grains.  The three components of the 
conductivity tensor are plotted in the lower panel.  $ 
\sigma_\parallel $ dominates the other two components until $ \nh\ga 
10^{15} \percc $.  Below about $ 10^{10} \percc $, the Hall term $ 
\sigma_1 $ is negative and is typically an order of magnitude below 
the Pedersen term, $ \sigma_2 $.  $ \sigma_1 $ changes sign at 
roughly $ 10^{10} \percc $.  Above this density, $ |\sigma_1 |$ 
approaches $ \sigma_2 $ and the ambipolar diffusion approximation 
breaks down for $ \nh \ga 10^{13}\percc $.

The contributions of different charged species to the components of $ 
\bmath{\sigma} $ are plotted in Figure \ref{fig:NU90_contributions}.  
It should be noted that each species contributes positively to both $ 
\sigma_\parallel $ and $ \sigma_2 $, but that the contribution to 
$ \sigma_1 $ carries the sign of $ Z_j $.  The conductivity 
parallel to the magnetic field is dominated by the electrons, even 
when much less abundant than the other charged species, because of 
their low rate coefficient for neutral scattering.  As one might 
expect, the grain contributions are negligible because of their large 
collision cross-section with the neutrals.  Negatively charged grains 
dominate the Hall term below $ 10^{10} \percc $ at which point the 
ion Hall parameter has dropped to order unity, the ion contribution 
takes over and $ \sigma_1 $ becomes positive.  At the highest 
densities, positive grains are also important.  The Pedersen 
conductivity $ \sigma_2 $ is determined by the ions, except at the 
very highest densities when electrons start to contribute.

\begin{figure}
\centerline{\epsfxsize=8cm \epsfbox{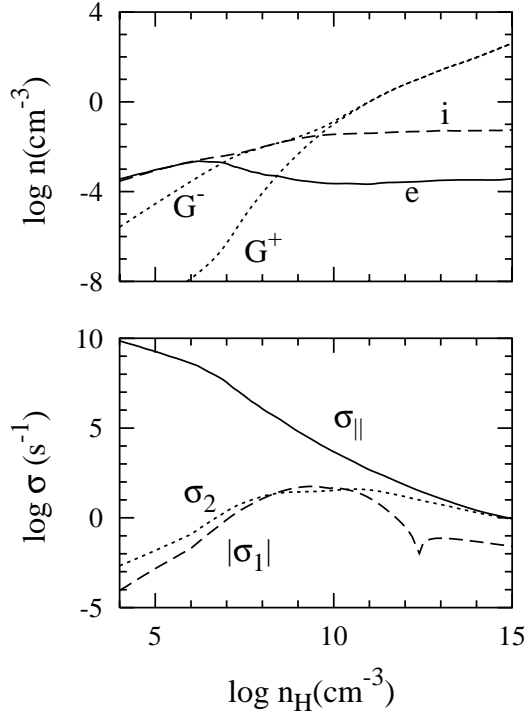}}\vskip 0cm
\caption{As for Fig. \ref{fig:NU90_conductivity}, but for an MRN 
grain-size distribution.} 
\label{fig:MRN_conductivity}  %\subsubsection{fig:MRN conductivity}
\end{figure}

\begin{figure}
\centerline{\epsfxsize=8cm \epsfbox{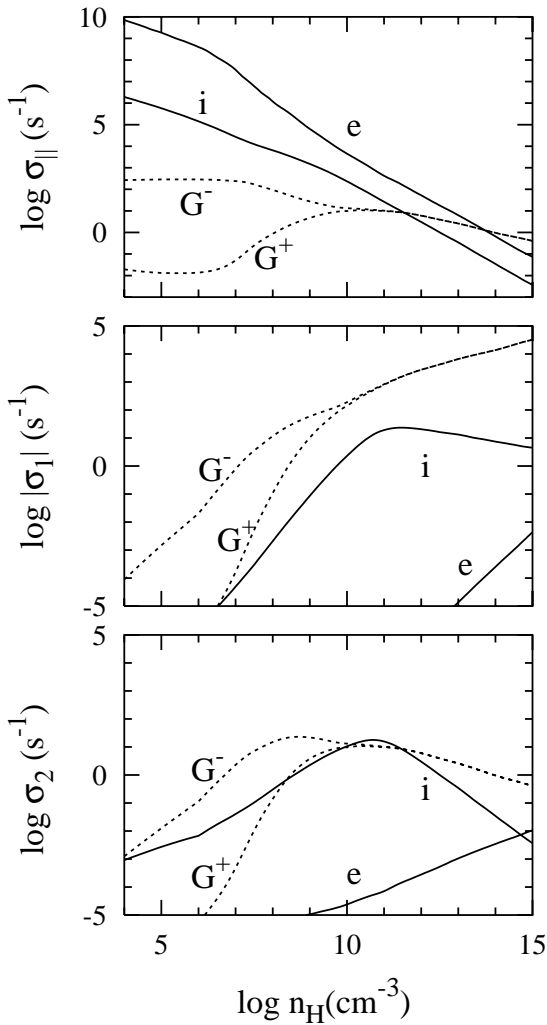}}\vskip 0cm
\caption{As for Fig. \ref{fig:NU90_contributions}, but for an MRN 
grain-size distribution.} 
\label{fig:MRN_contributions} %\subsubsection{fig:MRN contributions}
\end{figure}

The abundances and conductivity for the MRN model are plotted in 
Figure \ref{fig:MRN_conductivity}.  There are many more grains than in 
the single-size grain model, and grains therefore play a significant 
role in the ionisation balance at lower densities (Nishi et al 1991).  
Negatively charged grains dominate the electrons for $ \nh\ga 10^7 
\percc $, and positive grains dominate ions above $ 10^{10} \percc 
$.  The drop in electron no.  density reduces $ \sigma_\parallel $ 
considerably, but $ \sigma_\parallel $ is still much larger than the 
other components for $ \nh\la10^{12}\percc $. $ \sigma_1 $  
becomes comparable to $ \sigma_2 $ between $ 10^7 \percc $ and 
$ 10^{11} \percc $.  It drops sharply at higher densities, changing 
sign at $ 10^{12}\percc $.

Fig. \ref{fig:MRN_contributions} shows which species contribute to the 
components of $ \bsigma $.  Electrons again dominate $ \sigma_\parallel $,
but $ \sigma_1 $ and $ \sigma_2 $ are determined by the grains.  
Note that for $ \nh\ga10^{10}\percc $, the positive and negative 
grains contribute nearly-equal terms of opposite sign to the Hall 
conductivity.  The slightly depressed abundance of positive grains (of 
order 0.1--1\%) is sufficient for the net grain contribution to $ 
\sigma_1 $ to dominate $ \sigma_2 $.  Above $ 10^{11}\percc $ the 
asymmetry drops to the point that $ |\sigma_1| $ is a small fraction 
of $ \sigma_2 $, becoming dominated by the ion contribution above $ 
10^{12}\percc $.

\begin{figure}
\centerline{\epsfxsize=8cm \epsfbox{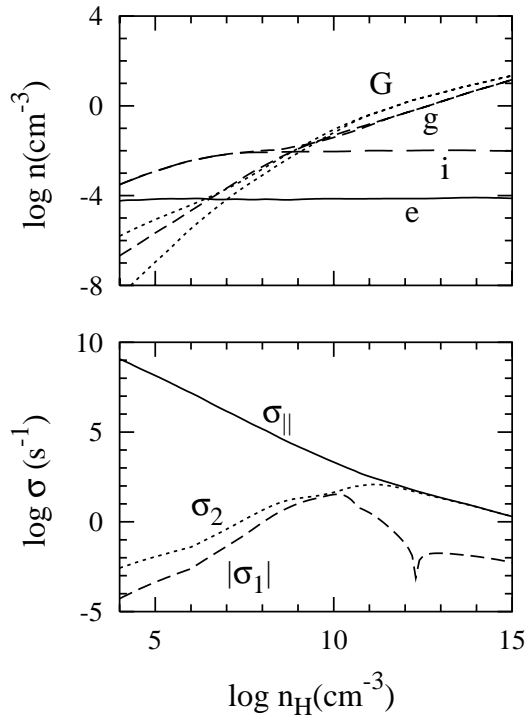}}\vskip 0cm
\caption {As for Fig. \ref{fig:NU90_conductivity}, but for an MRN 
grain-size distribution and PAHs} 
\label{fig:PAH_conductivity} %\subsubsection{fig:PAH conductivity}
\end{figure}
\begin{figure}
\centerline{\epsfxsize=8cm \epsfbox{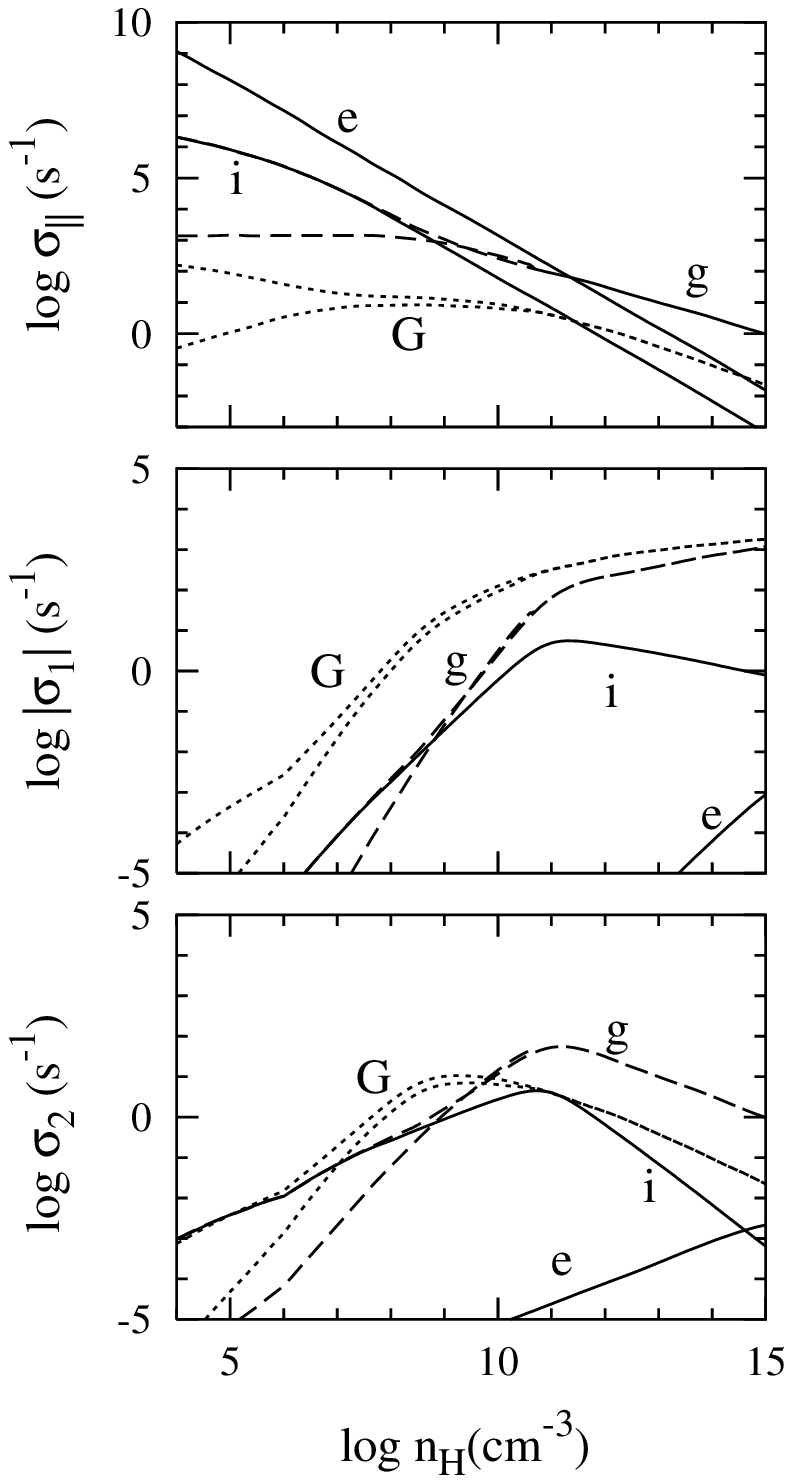}}\vskip 0cm
\caption {As for Fig. \ref{fig:NU90_contributions}, but for an MRN 
grain-size distribution and PAHs.} 
\label{fig:PAH_contributions} %\subsubsection{fig:PAH contributions}
\end{figure}

Finally, the effects of a PAH population are shown in Figures 
\ref{fig:PAH_conductivity} and \ref{fig:PAH_contributions}.  
Ions and PAHs are the dominant charged species between $ 10^6 
$--$10^9 \percc $.  For $ \nh \ga 10^9 \percc $ PAHs and grains 
of either sign are important.  Despite the change to the charged 
particle abundances, the components of the conductivity tensor behave 
similarly to the MRN-only case, with $ |\sigma_1|\approx \sigma_2 $ 
between $ 10^8 $ and $ 10^{11}\percc $ .  Electrons continue to dominate $ 
\sigma_\parallel $ below $ 10^{10}\percc $, with PAHs dominating at 
higher densities.  Grains determine $ \sigma_1 $ and $ \sigma_2 
$, with PAHs becoming important above $ 10^{10} \percc $.

\begin{figure} %\subsubsection{fig:B-n plane}
\centerline{\epsfxsize=8cm \epsfbox{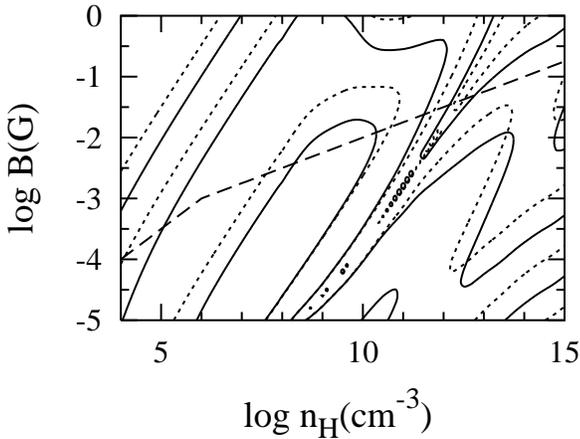}}\vskip 0cm
\caption {The ratio $ |\sigma_1|/\sigma_2 $ for temperatures of 30 
and 300 K (solid and dotted contours respectively).  The contours are 
spaced logarithmically, with levels 1, 0.1, and 0.01.  The 
dashed curve indicates the magnetic field scaling adopted in 
calculating Figs \ref{fig:NU90_conductivity} -- 
\ref{fig:PAH_contributions}.} 
\label{fig:B-n_plane} 
\end{figure}
A broad range of magnetic field strengths are appropriate to any 
particular value of the gas density, so it is important to consider 
the effects of departures of the magnetic field from the scaling 
(\ref{eq:B_scaling}).  Thus we show in Fig.  \ref{fig:B-n_plane} how 
the ratio $ |\sigma_1|/ \sigma_2 $ depends on both $ B $ and $ 
\nh $ for the MRN grain model.  Consider first the solid contours, 
which refer to a gas temperature of 30 K. The innermost contour 
corresponds to a ratio of unity, with successive contours indicating 
where $ |\sigma_1|/ \sigma_2 = 0.1$, and 0.01.  The gradients are 
weakest parallel to lines of constant $ B/\nh $ as the Hall 
parameters are constant and $ \bsigma $ varies only as a result of 
the density-dependence of the $ n_j $.The dashed trajectory 
indicates the scaling of $ B $ with $ \nh $ from eq.  
(\ref{eq:B_scaling}), thus Figure \ref{fig:MRN_conductivity} shows $ 
\bsigma $ for cuts taken along this trajectory.  It is clear that $ 
|\sigma_1|\ga \sigma_2 $ over a significant region at densities 
characteristic of dense cores.  In addition there is a region at 
higher densities ($ \nh\approx10^{11} $--$ 10^{13}\percc $) and mG 
field strengths in which the the ratio is greater than 0.1.  The 
dashed curves show the effect of increasing the temperature to $ 300 
$ K, which increases the electron and grain momentum-transfer rate 
coefficients and generally increases $ |\sigma_1|/\sigma_2 $ (note 
that the effects of the increased temperature on the charged-particle 
abundances have not been included).

\section{Alfv\'en Waves}
\label{sec:Alfven_waves}

The results of the previous section show that the Hall term is 
significant for molecular gas densities in the range $ 10^8 $--$ 
10^{11} \percc $.  As outlined in \S\ref{sec:formulation} this will 
have consequences for dynamical processes, particularly for the 
time-evolution of magnetic fields from an initial state.  By way of 
illustration, we consider here the propagation of Alfv\'en waves in 
weakly-ionised media.

Linear wave propagation in molecular clouds has previously been 
studied by Kulsrud \& Pearce (1969), who assumed a plasma comprised of 
neutrals, ions and electrons, and showed that short wavelengths are 
rapidly damped by ion-neutral friction.  Pilipp et al 
(1987) examined the influence of charged grains, assumed to be all of 
the same size.  Pilipp et al considered interactions between 
charged species and the coupling between the charged and neutral grain 
fluids that results from charging and neutralisation processes.   
Propagation at an oblique angle to the magnetic field
has been considered by Cramer \& Vladimirov (1997). 

Here we present a simpler treatment using the formulation of 
\S\ref{sec:formulation}, which neglects the inertia of the charged 
species, and thus is valid for wave frequencies $ \omega $ 
below the collision frequencies $ \gamma_j \rho $ 
of the charged species with the neutrals.  This also permits us to 
neglect the coupling between charged and neutral grains, which 
generally is only important at even higher frequencies (Pilipp et al 
1987).

We begin by estimating the frequency $ \omega_c $ at which flux 
freezing breaks down, and showing that $ \omega_c \ll \gamma\rho_j $.  
This guarantees that our treatment applies to the interesting regime 
$ \omega\sim \omega_c $.  For long-wavelength disturbances, the $ c 
\curl \E' $ term in the induction equation (\ref{eq:induction}) is 
negligible and the effects of finite conductivity are small: the 
dispersion relation for Alfv\'en waves is simply $ \omega \approx k 
v_A $.  At successively shorter wavelengths the $ c \curl \E' $ 
term becomes increasingly important, and the wave modes are strongly 
modified by the finite conductivity when $ \omega \approx 
(\omega/v_A)^2 c^2 / 4\pi\sigma_\perp $, i.e.  at a critical 
frequency
\begin{equation}
	\omega_c = \frac{B^2 \sigma_{\perp}}{\rho c^2} \,.
	\label{eq:omega_c}
\end{equation}
Substituting for $ \sigma_\perp $ using equations (\ref{eq:sigma_perp}) and 
(\ref{eq:Hall_parameter}), and noting that the norm of a sum 
of vectors is less than the sum of the norms, we obtain a limit 
on $ \omega_c $:
\begin{equation}
	\omega_c < \sum_{j} 
	\frac{\gamma_j\rho_j|\beta_j|}{\sqrt{1+\beta_j^2}} < 
	\sum_{j}\gamma_j\rho_j \,.
 	\label{eq:omega_c_limit}
\end{equation}
Strict equality is approached in the limit $ |\beta_j|\rightarrow 
\infty $, i.e.  when all of the charged species are tied to the 
magnetic field lines, in which case $ \omega_c $ is the collision frequency of 
the neutrals with any of the charged species.  More generally, $ 
\omega_c $ is less than this and, as $ \rho_j\ll\rho $, is clearly 
much less than $ \gamma\rho_j $ for any charged species $ j $.  
The smallest $ \gamma_j \rho $ is that for the largest (i.e.  $ 
0.1\micron $) grains.  In dense clouds, $ \gamma_G \rho_G \sim 
\gamma_i \rho_i $, thus $ \omega_c \sim \gamma_G\rho_G $.  As the 
mass in grains is of order a percent of the neutral gas, and only a 
fraction of the grains are charged, this implies that the dispersion 
relation we obtain below is valid for frequencies up to at least $ 
100\, \omega_c $.  At higher frequencies the 
neutral component does not play a significant dynamical role but 
merely provides a background drag that damps small-scale disturbances 
in the ionised component of the fluid.

\subsection{Linearisation}
\label{subsec:linearisation}

We linearise equations (\ref{eq:continuity}), 
(\ref{eq:momentum}), (\ref{eq:j_curlB}), (\ref{eq:efluid}), 
and (\ref{eq:jsigmaE}) about a homogeneous undisturbed state with $ 
\v=0 $, $ \J=\E'=0 $, and $ \B=B\x $.  Changes in the 
conductivity associated with the perturbations do not appear in the 
linearised equations, as $ \E' $ vanishes in the unperturbed state.  
Thus we need not make any assumptions about how the charged particle 
abundances respond to the perturbations -- all that is needed is the 
conductivity tensor of the unperturbed fluid.

Considering waves propagating parallel to $\B$ of the form $ \exp 
i(\omega t - kx) $, and discarding an additional quadratic factor 
that yields the familiar isothermal sound waves propagating parallel 
and antiparallel to the magnetic field, we find that $ \omega $ and 
$ k $ are related by the two dispersion relations
\begin{equation}
	\omega^2 \mp k^2v_A^2 \exp (\pm i\theta) \, \omega/\omega_c - k^2 v_A^2 = 0\,
	\label{eq:dispersion_relation}
\end{equation}
where
\begin{equation}
	\theta = \cos^{-1}\left(|\sigma_1| / \sigma_{\perp} \right) \,,
	\label{eq:theta}
\end{equation}
may range between 0 and $ \pi/2 $.  The two dispersion relations 
yield four roots; assuming that $ k>0 $ we are 
interested in the two roots with positive real part.  Setting
\begin{equation}
	b = \frac{kv_A}{2\omega_c}\exp (\pm i\theta)
	\label{eq:b}
\end{equation}
these two roots may be written
\begin{equation}
	\omega_1 = \left(\sqrt{1+b^2} +b \right) kv_A
	\label{eq:omega1}
\end{equation}\
and
\begin{equation}
	\omega_2 = \frac{k^2v_A^2}{\omega_1^*} \,,
	\label{eq:omega2}
\end{equation}
where $ \omega_1^* $ is the complex conjugate of $ \omega_1 $.  
Note that $ |\omega_2| < kv_A < |\omega_1| $.

The conductivity parallel to the field, $ \sigma_{\parallel} $ does 
not appear in the dispersion relation because $\E'_{\parallel} = 0 $ 
for these modes.  The conductivity perpendicular to $ \B $, $ 
\sigma_{\perp} $, determines $ \omega_c $ and sets the length 
scale at which the dispersion relation is modified from the ideal MHD 
case by the finite conductivity of the plasma, and the relative sizes 
of the Hall and Pedersen conductivities sets $ \theta $.

Solving the linearised equations for the perturbations, we find:
\begin{equation}
\delta B_x = 0, \,\,\delta B_y = \pm i\, \mathrm{sign}(\sigma_1) \delta B_z \,.
	\label{eq:delta_B}
\end{equation}
\begin{equation}
	\bmath{\delta J} = \pm \mathrm{sign}(\sigma_1)\frac{kc}{4\pi}\dB \,.
	\label{eq:deltaJ}
\end{equation}
\begin{equation}
	\frac{\dv}{v_A} = - \left(\frac{kv_A}{\omega}\right) \frac{\dB}{B}
	\label{eq:deltav}
\end{equation}
\begin{equation}
	\dE' = -i\, \mathrm{sign}(\sigma_1) \exp (\pm i\theta) \,
	\frac{kv_A}{\omega_c} \, \frac{v_A}{c} \, \dB \,.
	\label{eq:deltaEfluid}
\end{equation}
The perturbation to the electric field in the rest frame of the 
unperturbed fluid is
\begin{equation}
	\dE = \mp i\, \mathrm{sign}(\sigma_1) \left(\frac{\omega}{kv_A}\right) \frac{v_A}{c} \,\dB \,.
	\label{eq:deltaElab}
\end{equation}
Equation (\ref{eq:delta_B}) shows that the wave modes are circularly 
polarised, with the sense of polarisation depending 
on the sign of $ \sigma_1 $.

\subsection{Wave propagation}
\label{subsec:wave_propagation}

\begin{figure}
\centerline{\epsfxsize=8cm \epsfbox{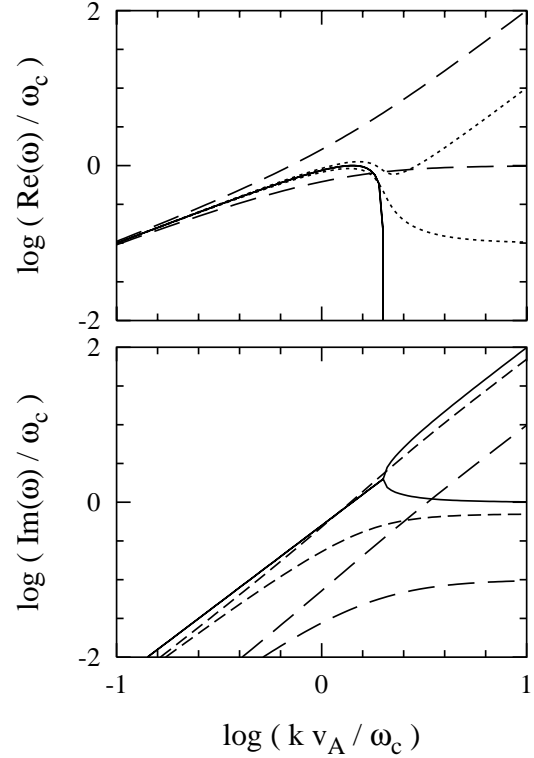}}
\vskip 0cm \caption{The dispersion relation for long-wavelength waves 
propagating parallel to the magnetic field in a uniform, 
weakly-ionised plasma.  The real and imaginary parts of the wave 
frequency $ \omega $ (upper and lower panels respectively) are 
plotted as a function of (real) wave number, $ k $.  The 
characteristic frequency $ \omega_c $ is defined in eq.  
(\ref{eq:omega_c}).  The curves are distinguished by the ratio of the 
Hall and Pedersen conductivities, $ |\sigma_1|/\sigma_2 $: 
\emph{solid} -- 0; \emph{dotted} -- 0.1; \emph{short-dashed} -- 1; 
\emph{long-dashed} -- 10. For clarity, the curves for the real part 
of $ \omega $ for $ |\sigma_1|/\sigma_2 = 1$, and for the 
imaginary part for $ |\sigma_1|/\sigma_2 =0.1 $ have been omitted as 
they are similar to those for $ |\sigma_1|/\sigma_2 =10 $ and 0 
respectively.}
\label{fig:dispersion_reln} %\subsubsection{fig:dispersion relation}
\end{figure}

The wave frequency $ \omega $ is plotted as a function of $ k $ in 
Fig.  \ref{fig:dispersion_reln} for different choices of $ 
|\sigma_1|/\sigma_2 $.  The solid curves give the real and imaginary 
parts of $ \omega $ for $ \sigma_1 = 0 $, the pure ambipolar 
diffusion case (see Kulsrud \& Pearce 1969).  For $ kv_A/\omega_c < 2 
$ the two circularly-polarised modes have the same frequency, thus one can construct 
linearly-polarised Alfv\'en waves.  At low wave numbers, the modes 
reduce to Alfv\'en waves in the combined neutral and ionised fluid.  
As the wave number is increased the modes propagate at slightly below 
the Alfv\'en speed and are damped, the damping rate increasing as $ 
k^2 $.  The phase speed drops significantly below the Alfv\'en speed 
at $ kv_A \approx \omega_c $.  At $ kv_A/\omega_c = 2 $ the waves 
become evanescent, with two distinct modes.

The other curves plotted in Fig. \ref{fig:dispersion_reln} show how 
the modes are modified as the Hall conductivity is increased in 
importance.  In general the damping rate is decreased because the 
Hall current is perpendicular to $ \E' $ and therefore does not 
contribute to the energy dissipation in the fluid.  When the Hall 
conductivity is non-zero, the evanescent regime disappears.

The two circularly-polarised modes propagate with different phase 
speeds and damping rates because the Hall conductivity introduces a 
handedness into the fluid -- the microphysical asymmetry between the 
positively- and negatively-charged species is manifested as a 
dependence on the sign of the magnetic field, or the sense of circular 
polarisation.

For $ kv_A/\omega_c \gg 1 $, the roots of the dispersion relations with 
positive real part are
\begin{equation}
	\frac{\omega_1}{\omega_c} \approx 
	\left(\frac{kv_A}{\omega_c}\right)^2 \exp (i\theta)
	\label{eq:omega_1_highfreq}
\end{equation}
and 
\begin{equation}
	\frac{\omega_2}{\omega_c} \approx \exp (i\theta)
	\label{eq:omega_2_highfreq}
\end{equation}
A comparison of terms in the linearised induction equation yields
\begin{equation}
	\left|\frac{\partial \B / \partial t}{\curl (\v\cross\B) }\right| = 
	\left|\frac{\omega}{kv_A}\right|^2 \,.
	\label{eq:induction_terms}
\end{equation}
For $ \omega_1 $ or $ \omega_2 $ this ratio is $ 
(kv_A/\omega_c)^2 $ or $ (kv_A/\omega_c)^{-2} $ respectively.  Thus 
$ \curl\E' $ in the induction equation is balanced by $ 
\partial \B / \partial t $ and $ \curl (\v\cross\B) $ respectively.  
Equation (\ref{eq:deltav}) shows that for these two modes, the ratio 
$ |\dB/B| / |\dv/v_A|$ is $ (kv_A/\omega_c) $ or $ 
(kv_A/\omega_c)^{-1} $.  The highly-damped mode $ \omega_1 $ mode 
can be regarded as a wave in the \emph{ionised} component fluid that 
is damped by collisions with the nearly static neutrals.  The 
$ \omega_2 $ mode corresponds to transverse oscillations in the 
neutrals with small perturbations in the magnetic field and ionised 
component (c.f.  Kulsrud \& Pearce 1969).  Note that both modes are 
undamped in the limit that $ \sigma_2 \rightarrow 0 $.

In the low-frequency limit $ kv_A/\omega_c \ll 1$, $ \omega_1 $ and $ 
\omega_2 $ become, to first order in $ kv_A/\omega_c $,
\begin{equation}
	\frac{\omega}{k v_A} \approx 1 \pm \frac{kv_A}{2\omega_c} \cos\theta
	 - i \frac{kv_A}{2\omega_c}\sin\theta
	\label{eq:omega_lowfreq}
\end{equation}
Thus the real part of $ \omega $ (i.e.  phase speed) is hardly 
affected by $ \bsigma $, but the damping rate is dependent on both 
$ \sigma_1 $ and $ \sigma_2 $.  It should be noted that including 
the Hall term decreases the damping rate for a given $ k $ by 
increasing $ \omega_c $ and introducing a factor of $ \sin\theta 
$.  Thus the damping rate is reduced by a factor of $ \sin^2\theta 
$ once $ \sigma_1 $ is taken into account.

\section{Discussion}
\label{sec:discussion}

The evaluation of the conductivity tensor presented in 
\S\ref{sec:ionisation} shows that for an MRN grain-size distribution 
the Hall conductivity is important for densities between $ 10^7 $ 
and $ 10^{11} \percc $.  PAHs, if present, reduce this range to $ 
10^9 $--$ 10^{10}\percc $.  The magnitude of $ \sigma_1 $ is 
sensitive to the slight difference (1\% or less) in the number
densities of positive and negative grains, and positive and negative 
PAHs, so a careful calculation of the ionisation balance at these 
densities is required to confirm these results.  The sensitivity of 
the conductivity tensor to small changes in the relative abundances of 
these species potentially couples the qualitative dynamical 
behaviour of dense molecular gas to chemical reactions that affect the 
ionisation level in the gas, as these in turn may depend on the 
relative drift of the charged species and neutrals.

In dense molecular clouds the grain-size distribution is modified by 
the agglomeration of small grains through thermal collisions and the 
sweeping up of smaller particles by large particles (Ossenkopf 1993; 
Ossenkopf \& Henning 1994).  These processes remove most of the grains 
that have Hall parameters of order unity (grain sizes of order 100 
\AA ), which tends to reduce the size of $|\sigma_1|$.  In this 
case one expects the conductivity to resemble that for the $0.1$ 
micron grain case (see Figs 1 and 2).

At a given density and magnetic field strength, the conductivity also 
depends on the assumed cosmic-ray ionisation rate per unit volume, $ 
\zeta \nh $.  There is a simple scaling law for $ \bsigma $ as the 
fractional abundances $ n_j/\nh $ of the charged species depend only 
on the ratio $ \nh/\zeta $ (e.g.  Nishi et al 1991).  This allows $ 
\bsigma $ to be written as a function of $ \nh/\zeta $ and $ 
B/\zeta $ rather than of $ \nh $, $ B $ and $ \zeta $ 
independently.  Thus, for example, if $ \zeta $ is increased from 
our adopted value of $ 10^{-17} \ut H -1 \persec $ to $ 10^{-16} 
\ut H -1 \persec $, the contours in Fig.  \ref{fig:B-n_plane} are 
shifted upwards and to the right by 1 logarithmic unit in each 
direction, significantly increasing the area covered by the $ 
|\sigma_1| = \sigma_2 $ contour.  The cosmic-ray flux is 
substantially reduced for densities in excess of $ 10^{13}\percc $, 
which occur near the midplane in the inner few AU of protostellar 
disks, because the cosmic-ray flux is substantially reduced when the 
disk surface density exceeds the cosmic-ray attenuation column, $ 
96\u g \ut cm -2 $. In particular, the surface density at 1 AU from 
the Sun in the minimum solar nebula is approximately 1200 $ \u g \ut 
cm -2 $ (Hayashi 1981).  Under these conditions the charged particle 
abundances are substantially reduced and the Hall contribution is 
negligible.

The Hall term introduces a definite handedness to the dynamical 
behaviour of the gas because of the differences in the properties of 
negatively- and positively-charged species.  There is little change in 
dynamics on the largest scales, for which flux-freezing is a good 
approximation.  On small length scales, the dynamical behaviour of the 
fluid is qualitatively altered, and is no longer unaffected under a 
global reversal of magnetic field direction.  This has been previously 
noted in models for the acceleration of outflows from protostellar 
disks (Wardle \& K\"onigl 1993), where the relative orientation (i.e.  
parallel or anti-parallel) of the large-scale threading the disk and 
the angular velocity of the disk affects the ability of the field to 
centrifugally drive a disk wind.  These effects are also apparent in 
models of C-type shock waves (Wardle 1997).

We chose to illustrate these effects by examining the consequences for 
Alfv\'en wave propagation, as the wave modes
encapsulate the dynamical response of the fluid to perturbations.  
On large length scales, the Alfv\'en modes are damped more rapidly if 
the Hall contribution to $ \bsigma $ is included, which could in 
principle have consequences for the survival of the Alfv\'en waves 
believed to support molecular clouds parallel to the magnetic field 
(Arons \& Max 1975; Fatuzzo \& Adams 1993; Gammie \& Ostriker 1996), 
and for the contribution of wave dissipation to cloud heating (Zweibel 
\& Josafatsson 1983).  However at the relevant densities $ |\sigma_1| 
\ll \sigma_2 $, and in any case the factor of two difference that 
would obtain when $ |\sigma_1| \approx \sigma_2 $ is of the same 
order as the theoretical and observational uncertainties in the field 
strength and charged particle abundances that determine the damping 
rate.  Of more importance are the changes that occur on the length 
scales on which flux-freezing breaks down, when the two 
circularly-polarised Alfv\'en modes have different phase speeds and 
damping rates.  For example, consider the following thought experiment 
that follows from the linear analysis of \S\ref{subsec:linearisation}.  
Imagine at some initial time a homogeneous, static fluid permeated by 
a magnetic field of the form $ \B=B_0\x + B_1\sin kx \, \bmath{y} $, 
where $ B_0 $ and $ B_1 \ll B_0 $ are positive constants.  How 
will the system evolve with time?  In the ambipolar diffusion limit $ 
B_y $ will oscillate back and forth with nodes where $ \sin kx = 0 
$, the oscillatory motion being damped by ambipolar diffusion.  If 
the Hall term is important the motion is more complicated as one of 
the circular polarisations is more rapidly damped than the other.  The 
field tends to twist back and forth, with an associated sloshing of 
the fluid.  Thus the evolution is quite different in the two cases.

As the transport of angular momentum by magnetic fields and the 
breakdown of flux-freezing are believed to play important roles in 
the collapse of cloud cores to form stars (e.g.  Shu, Adams \& Lizano 
1987; Mouschovias 1987), 
and the densities pass through the regime in which the Hall term is 
important, the consequences for the dynamics of collapsing cloud cores 
may be profound.  Norman \& Heyvaerts (1985) also noted that the Hall 
component of the conductivity tensor may become significant during 
star formation.  However they dismiss its effects, claiming that the 
Hall current would rapidly lead to a slight charge separation, with 
the associated electric field quenching the Hall current.  Although 
this can certainly occur for particular idealised geometries with 
well-defined boundaries, it is not clear that this mitigation of the 
Hall effect is generally true.  For example, consider the quasistatic 
collapse of an axisymmetric cloud core through a magnetic field, and 
suppose that at some initial time the core is not rotating and the 
magnetic field is poloidal.  Then the current is toroidal and the 
electric field in the neutral fluid frame is poloidal.  Thus $ 
\grad\cross\E $, hence $ \partial\B/\partial t $, is poloidal, as 
is the Lorentz force $ \J\cross\B/c $.  Thus the fluid velocities 
and the magnetic field remain poloidal as the system evolves.  In 
contrast, if the Hall term is important, at $ t=0 $ the electric 
field has a poloidal component (as $ \J $ is still toroidal 
initially) and $ B_{\phi} $ will grow with time, forcing the fluid 
to also pick up a toroidal component.  It is not hard to show that if 
the Hall and ambipolar diffusion components of the conductivity are of 
the same order that the toroidal components of the velocity and 
magnetic field will contribute significantly to the gravitational 
support of the core and therefore will significantly affect the 
dynamics of gravitational collapse.  In this example the boundaries on 
which the charge is presumed to build up initially would have to be 
surfaces of constant azimuth, which cannot exist under the assumption 
of axisymmetry.

\section{Summary}
\label{sec:summary}
In this paper we evaluated the conductivity tensor for weakly-ionised, dense molecular 
gas assuming a temperature of 30 K and a cosmic-ray ionisation rate of 
$ 10^{-17}\ut s -1 \ut H -1 $. We also considered the propagation of 
Alfv\'en waves in this medium.  Our results are 
summarised as follows:
 \begin{enumerate}
 	\item  For single-size grains of size 0.1 $ \micron $, the Hall term 
 	does not become significant.
 
 	\item  For an MRN grain-size distribution
 	the Hall term becomes important between $ 10^7 $ 
	and $ 10^{11} \percc $.  
	
	\item PAHs, if present, reduce this range to $ 10^9 $--$ 
	10^{10}\percc $.
	
	\item The magnitude of the Hall conductivity is sensitive to the 
	small differences in abundances of positive and negative grains, and 
	positive and negative PAHs.
 
 	\item  Long-wavelength Alfv\'en waves, for which flux freezing 
 	almost holds, are not much affected by the 
 	Hall term, apart from a reduced damping rate.  Generally this 
 	reduction is not significant given the uncertainties in the 
 	parameters of the gas in molecular clouds.
 
	\item At wavelengths comparable to or less than the length scale 
	on which flux-freezing breaks down, left- and right-circularly 
	polarised Alfv\'en waves of the same frequency have distinct phase 
	speeds and damping rates.  For the typical case, the right-hand 
	(left-hand) polarisation is most strongly damped for propagation 
	parallel (anti-parallel) to the magnetic field.  The other mode is 
	damped on the ion-neutral coupling time scale, even at the 
	shortest wavelengths.
  \end{enumerate}
We noted that the modification to Alfv\'en wave propagation is 
indicative of the change in the dynamical behaviour of the gas.  In 
particular we suggested that the dynamics of the collapse of dense 
cloud cores may be profoundly affected.

We thank Leon Mestel for useful discussions.  C. Ng was supported by 
an SRCfTA vacation scholarship.  The Special Research Centre for 
Theoretical Astrophysics is funded by the Australian Research Council 
under the Special Research Centres programme.

\bsp
\label{lastpage}
\end{document}